\documentclass[aps,prl,superscriptaddress,floatfix,10pt,twocolumn]{revtex4-2}

\usepackage[utf8]{inputenc}
\usepackage[T1]{fontenc}
\usepackage[main=english]{babel}
\usepackage{mathtools}
\usepackage{amssymb}
\usepackage{bm}
\usepackage{mathrsfs}

\usepackage{siunitx}
\usepackage{braket}

\usepackage{graphicx}
\usepackage{overpic}
\usepackage[dvipsnames]{xcolor}
\usepackage[
colorlinks=true,allcolors=blue,%
pdfusetitle,%
]{hyperref}

\AtBeginDocument{
\newcommand{\dif}{\mathrm{d}}
\newcommand{\E}{\mathrm{e}}
\newcommand{\I}{\mathrm{i}}
\newcommand{\PI}{\pi}
 \let\Im\relax

\DeclareMathOperator{\Im}{Im}

}

\usepackage{microtype}

\begin{document}
\title{Polariton-induced superconductivity in two-dimensional metals}
\author{Riccardo Riolo}
\affiliation{Dipartimento di Fisica dell'Universit\`a di Pisa, Largo Bruno Pontecorvo 3, I-56127 Pisa,~Italy}
\author{Frank H.L. Koppens}
\affiliation{ICFO-Institut de Ci\`{e}ncies Fot\`{o}niques, The Barcelona Institute of Science and Technology, Av. Carl Friedrich Gauss 3, 08860 Castelldefels (Barcelona),~Spain}
\affiliation{ICREA-Institució Catalana de Recerca i Estudis Avançats, Barcelona, Spain}
\author{Pablo Jarillo-Herrero}
\affiliation{Department of Physics, Massachusetts Institute of Technology, Cambridge, Massachusetts,~USA}
\author{Giacomo Mazza}
\affiliation{Dipartimento di Fisica dell'Universit\`a di Pisa, Largo Bruno Pontecorvo 3, I-56127 Pisa,~Italy}
\author{Allan H. MacDonald}
\affiliation{Department of Physics, The University of Texas at Austin, Austin, TX 78712,~USA}
\author{Marco Polini}
\affiliation{Dipartimento di Fisica dell'Universit\`a di Pisa, Largo Bruno Pontecorvo 3, I-56127 Pisa,~Italy}
\affiliation{ICFO-Institut de Ci\`{e}ncies Fot\`{o}niques, The Barcelona Institute of Science and Technology, Av. Carl Friedrich Gauss 3, 08860 Castelldefels (Barcelona),~Spain}
\begin{abstract}
The electronic properties of two-dimensional (2D) metals are altered by 
changes in their three-dimensional dielectric environment.  
In this Letter we propose that superconductivity can be induced in a 
2D metal by resonant coupling between its plasmonic 
collective modes and optical phonons in a nearby polar dielectric.  
Specifically, we predict that relatively high-temperature superconductivity 
can be induced in bilayer graphene twisted to an angle somewhat 
larger than the magic value by surrounding it with a THz polar dielectric. Our conclusions are based on numerical solutions of Eliashberg equations
for massless Dirac fermions with tunable Fermi velocities and Fermi energies, 
and can be understood qualitatively in terms of a generalized McMillan formula.
\end{abstract}
\maketitle

\noindent{\color{blue}\it Introduction}.---This Letter proposes a cavity-based
strategy to achieve a feat that is often considered to be the holy grail of quantum 
materials engineering, inducing superconductivity in an otherwise
normal metal.  Cavity materials engineering~\cite{lu_advancedphotonics_2025,hubener_mqt_2024,ebbesen_chemrev_2023,bloch_Nature_2022,schlawin_APR_2022,garciavidal_Science_2021,genet_PT_2021,rubio_NatureMater_2021} is a growing research field whose broad aim is to modify the 
properties of quantum matter (in a tunable fashion) by coupling electronic degrees of 
freedom to a suitably engineered electromagnetic environment.  
Electromagnetic radiation is regularly used to alter properties by 
driving matter into non-equilibrium excited states.  For example, Persky et al.~\cite{persky_arXiv_2025} have recently 
established a dependence of orbital magnetism in magic angle twisted bilayer graphene (on ${\rm WSe}_2$) on electromagnetic radiation in the near-infrared spectral range. Evidence of light-induced non-equilibrium superconducting phases 
has been reported in cuprates~\cite{fausti_science_2011,mankowski_nature_2014,vonhoegen_prx_2022,ribak_prb_2023} 
and organic 
superconductors~\cite{mitrano_nature_2016,cantaluppi_natphys_2018,buzzi_prx_2020,budden_natphys_2021,rowe_natphys_2023}.
Importantly, Fava et al. have observed transient magnetic field expulsion ~\cite{fava_nature_2024} corroborating the superconducting nature of these non-equilibrium states of matter. Theory predicts~\cite{sentef_scienceadvances_2018,schlawin_prl_2019,curtis_prl_2019,chakraborty_prl_2021,lu_pnas_2024,kozin_prb_2025} that permanent changes in the superconducting properties of a given material, including its critical temperature $T_{\rm c}$,
can also be achieved in equilibrium in the presence of a suitably engineered electromagnetic environment.
Passive control of superconducting properties is more powerful than 
transient control in systems driven from equilibrium and 
is therefore a principle aspiration of cavity materials engineering. 

\begin{figure}[t]
\centering
\includegraphics{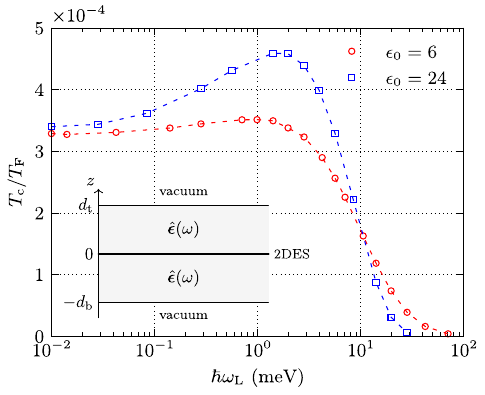}
\caption{Polariton enhancement of superconductivity in a MDF 2DES.
The inset shows a sketch of the device analyzed in this work. 
The MDF 2DES is encapsulated between two slabs of polar material with 
thicknesses $d_{\rm t}$ and $d_{\rm b}$. In the main panel we plot the ratio $T_{\rm c}/ T_{{\rm F}}$, where $T_{\rm F}$ is the MDF 2DES Fermi temperature at velocity $v$,
as a function of the longitudinal optical phonon frequency $\hbar\omega_{\rm L}$. The results in this plot have been obtained by setting $d_{\text{t}}=d_{\text{b}}\to\infty$, $\epsilon_\infty=3$, carrier density $n = \SI{e11}{cm^{-2}}$, and Fermi velocity $v=v_0/2$ ($\theta=2.2^\circ$ and $T_{\rm F} = 150~{\rm K}$ in TBG). The 
two curves refer to different values of the phonon oscillator strength $\zeta \in [0,1]$ and hence to different values of $\epsilon_0$.  We argue in the main text
that these $T_{\rm c}$'s estimates are most reliable near the curve maxima.}
\label{fig:one}
\end{figure}

The impact of cavity control in equilibrium is usually weak 
because the dimensionless parameter of light-matter interactions, the fine-structure constant $\alpha = e^2/\hbar c \simeq 1/137$, is small. 
It is, however, well known that confining electromagnetic fields to small regions of space by using sub-wavelength cavities (plasmonic ~\cite{hugall_acsphoton_2018,bitton_acr_2022} or polaritonic~\cite{Basov_Science_2016,Low_NatMater_2017,Zhang_Nature_2021,Basov_Nanophotonics_2021,Plantey_ACS_2021}) can increase coupling between matter degrees-of-freedom and bosonic modes with large spectral densities-of-states. For example, the coupling between electronic excitations and polaritons
can be ultra-strong.  
One example that has been studied in great detail with near-field optical spectroscopy~\cite{woessner_naturemater_2015,lundeberg_Science_2017,alonso_NatureNano_2017,ni_nature_2018}, is the coupling between plasmons in two-dimensional (2D) 
materials such as graphene and phonon polaritons in a nearby (hyperbolic) dielectric, such as hexagonal boron nitride (hBN)~\cite{dai_science_2014,caldwell_naturecommun_2014,li_NatCommun_2015,dai_NatCommun_2015}.  
Indeed, this Letter builds on a theory of
electron-electron interactions in materials embedded in sub-wavelength/polaritonic cavities recently formulated in Refs.~\cite{andolina_prb_2024,Riolo_pnas_2025} and applied to the Shubnikov-de Haas~\cite{Riolo_pnas_2025} and quantum Hall regimes~\cite{andolina_arxiv_2025}.

After years of theoretical anticipation~\cite{lu_advancedphotonics_2025,hubener_mqt_2024,ebbesen_chemrev_2023,bloch_Nature_2022,schlawin_APR_2022,garciavidal_Science_2021,genet_PT_2021,rubio_NatureMater_2021}, experimental results proving the feasibility of cavity materials engineering are finally beginning to appear. Faist and collaborators~\cite{paravicini_bagliani_NaturePhys_2019,appugliese_science_2022,enkner_prx_2024,enkner_nature_2025} have shown that the transport properties of 2D electron systems in GaAs/AlGaAs quantum wells, in the integer and fractional quantum Hall regime, can be modified by sub-wavelegth cavities (metallic split-ring resonators) in the Terahertz (THz) spectral range. Jarc et al.~\cite{jarc_nature_2023} have 
demonstrated that the metal-to-insulator
transition in transition metal dichalcogenides embedded in low-energy THz cavities can be reversibly controlled by the cavity geometry. Very recently, Keren et al.~\cite{keren_arXiv_2025} have presented experimental data demonstrating
the modification of the superfluid density of a thick slab of a molecular superconductor ($\kappa$-ET) by the phonon polariton modes of $20$-$100$~{\rm nm} 
slabs of hBN. The key idea of Ref.~\cite{keren_arXiv_2025} is that the phonon mode which is responsible for superconductivity in $\kappa$-ET falls in the range of the hyperbolic phonon polariton modes of hBN. This leads to mode hybridization between $\kappa$-ET and hBN, which in turns leads to a transfer of spectral weight and the subsequent modification of the superfluid density of the phonon superconductor. In this case, the superfluid density (and correspondingly $T_{\rm c}$) of the resonant heterostructure hBN/$\kappa$-ET is {\it suppressed} with respect to that  isolated $\kappa$-ET.  In this Letter we show that the superconducting critical temperature $T_{\rm c}$ of a system can instead be {\it increased} by polaritons when they are resonant with a 2D metal's plasmonic collective modes.

\noindent{\color{blue}\it Device geometry and theoretical modelling}.---We consider a massless Dirac fermion (MDF) 2D electron system (2DES) with $N_{\text{f}}$ fermion flavors and tunable Fermi velocity $v$.  
MDFs with tunable $v$ can be realized in twisted bilayer graphene (TBG)~\cite{lopes_prl_2007,shallcross_prl_2008,mele_prb_2010,li_naturephys_2010,shallcross_prb_2010,bistritzer_prb_2010,lopes_prb_2012,morell_prb_2010,cao_prl_2016}, which has symmetry-protected Dirac cones.  (With the development of the quantum twisting 
microscope~\cite{inbar_nature_2023}, $v$ can now be varied {\it in situ}.) The relevant interaction in this material is the repulsive instantaneous Coulomb interaction $V(q,\infty)\equiv \lim_{\omega \to \infty} V(q,\omega)$, where $V(q,\omega)$ is defined below in Eq.~(\ref{eq:dressed_propagator_non_hyperbolic}).
Recent experiments~\cite{Barrier_arxiv2024,Gao_arxiv2024} on superconductivity in TBG suggest that electron-electron interactions play a major role in the superconducting instability of this intriguing material.
It has long been recognized that when dynamical screening ~\cite{Pines_and_Nozieres,Mahan_2000,Giuliani_and_Vignale} is 
taken into account but vertex corrections are neglected, repulsive Coulomb interactions
in metals on their own 
produce an instability towards a plasmon-mediated superconducting state~\cite{Takada_1978,Takada_1992a,Grabowski_1984,Sham_1985,Vignale_prb_1989,intVeld_2023} with a small but finite critical temperature, which we dub $T_{{\rm c}, \infty}$ below.
(We return to the validity of neglecting vertex corrections later.) We show here that when the plasmon modes~\cite{grigorenko_review_plasmonics} of a  
MDF 2DES are resonantly coupled to the longitudinal optical phonon modes of a nearby polar material, the same theory predicts critical temperature $T_{\rm c}$ that is higher than $T_{{\rm c}, \infty}$ and under better theoretical control.
\begin{figure}
\centering
\includegraphics{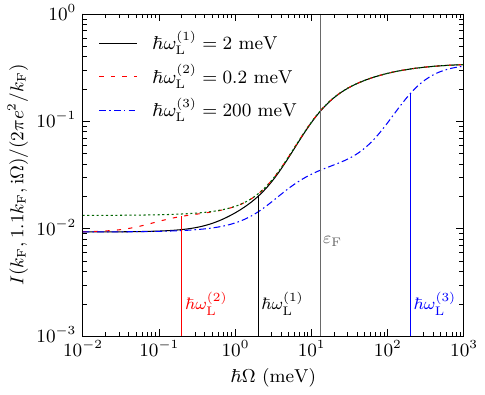}
\caption{Resonantly enhanced electron-electron interactions. The $s$-wave (i.e.~$\ell=0$) averaged interaction $I(k_{\text{F}},1.1k_{\text{F}},\I \Omega)$, defined in Eq.~(\ref{eq:avg-int-l}), is plotted as a function of the frequency $\Omega$ on the imaginary-frequency axis.
Results in this figure have been calculated for the same parameters as in Figs.~\ref{fig:one}. 
In particular, $\epsilon_\infty=3$, $\epsilon_0=24$,
and the energy of the longitudinal optical phonon mode
$\hbar\omega_{\text{L}}$ is reported in the legend.
The thin vertical lines are guides to the eye, representing the energy scales
discussed in the main text,
$\hbar\omega_{\text{L}}=\SI{2}{meV}$, $\SI{0.2}{meV}$, $\SI{200}{meV}$,
and the Fermi energy $\varepsilon_{\text{F}}$. The (green) dotted line is the averaged interaction in the absence of phonons, i.e.~calculated with permittivity $\epsilon(\omega)\equiv\epsilon_\infty$.}\label{fig:two}
\end{figure}

The device we have in mind is represented in Fig.~\ref{fig:one}. 
In the TBG platform, the number of fermion flavors is $N_{\text{f}}=8$ and
the MDF velocity is given approximately by~\cite{bistritzer_pnas_2011},
\begin{equation}\label{eqn:slow_velocity}
v = v_0 \frac{1 - 3\alpha^2(\theta)}{1+6\alpha^2(\theta)}~,
\end{equation}
where $v_0$ is the Fermi velocity in single-layer graphene~\cite{castroNeto_RMP_2009}, 
$\theta$ is the twist angle, and $\alpha(\theta) = \alpha_0\sin\left(\frac{\theta}{2}\right)^{-1}$, $\alpha_0$ being a constant that depends on microscopic details of TBG. Crucially, $v$ can be reduced to arbitrarily small values by tuning the twist 
angle towards the first magic angle $\theta^\star$~\cite{bistritzer_pnas_2011}. In the following, we fix $\alpha_0 = 5.5\times10^{-3}$, corresponding to  $\theta^\star\sim1.1^\circ$. 
In this work, we focus on $\theta>\theta^\star$ since our Eliashberg theory of superconductivity (see below) relies on a translationally- and rotationally-invariant MDF continuum model and we wish to avoid the complications of very strong electronic correlations.   This simplification justifies the generalized McMillan approach~\cite{McMillan_1968} described in Ref.~\cite{SM}. We leave an analysis of the impact of dynamical screening and polaritons on superconductivity in magic angle  TBG~\cite{Cao_Nature_2018} at $\theta = \theta^\star$ for future work~\cite{Comment_Singapore}.

We now proceed by encapsulating the MDF 2DES between two homogeneous dielectric slabs of a polar material. For the sake of simplicity, we take the two slabs to be of the same thicknesses $d$ and assume that no metal gates are present. Including the latter is straightforward, and their role is merely quantitative: they simply tend to suppress the superconducting critical temperature with respect to the case when they are absent (see Fig.~\ref{fig:tcphonond} in Ref.~\cite{SM}). The dielectric permittivity $\epsilon(\omega)$ of the polar material is taken to be of the simplest possible form~\cite{Dielectric_Losses}:

\begin{equation}\label{eq:simple_permittivity}
\frac{1}{\epsilon(\omega)}
= \frac{1}{\epsilon_\infty} \,
\frac{\omega^2 - \omega_{\text{T}}^2}{\omega^2 - \omega_{\text{L}}^2}~,
\end{equation}
where $\omega_{\text{L}}$ ($\omega_{\text{T}}$) is the frequency of the longitudinal (transverse) optical phonon, $\epsilon_\infty$ the high-frequency permittivity, and $\epsilon_0$ the static permittivity. The polar material has a reststrahlen band for $\omega_{\rm T} < \omega < \omega_{\rm L}$, where  $\omega_{\text{T}}$ is related to $\omega_{\text{L}}$ by
the Lyddane-Sachs-Teller (LST) relation,
\begin{equation}\label{eq:LST_relation}
\frac{\omega_{\text{L}}^2}{\omega_{\text{T}}^2}
= \frac{\epsilon_0}{\epsilon_{\infty}}~.
\end{equation}
For future purposes, we introduce the following dimensionless quantity,
\begin{equation}
\zeta \equiv \sqrt{1 - \frac{\epsilon_\infty}{\epsilon_0}} \in [0,1]~,
\end{equation}
which physically represents the oscillator strength of the phonon mode. 

Using the approach of Ref.~\cite{Riolo_pnas_2025}, we find that the electron-electron interaction $V(q,\omega)$ in the 2DES is repulsive and frequency-dependent \cite{Hyperbolicity}:
\begin{equation}\label{eq:dressed_propagator_non_hyperbolic}
V(q,\omega) = \frac{2\pi e^2}{
\epsilon(\omega)q}F(q,\omega)
\end{equation}
where the form factor
\begin{equation}\label{eq:form_factor_non_hyperbolic}
F(q,\omega)  \equiv
\frac{\tanh(qd) + \epsilon(\omega)}{1 + \epsilon(\omega)\tanh(q d)}~.
\end{equation}
(Note that 
$\lim_{qd \to \infty} F(q,\omega)=1$. Numerical results in the main text have been obtained in this limit. Results for finite values of $d$ can be found in Ref.~\cite{SM}.) The frequency-dependence of $V(q,\omega)$ captures the long-range 
coupling between carriers in the 2DES and longitudinal optical phonons
in the nearby polar dielectrics, which introduces retardation on the $\hbar \omega_{\rm L}$ energy scale. The frequency domain structure of the electron-electron interaction, induced by the polar materials in which $\epsilon_0 > \epsilon_\infty$, plays the key role in what follows.  The property that 
$V(q,\omega)$ is less repulsive at low frequencies than at high frequencies (on the imaginary frequency axis, see below) opens up a McMillan window of effective attraction between electrons, which can lead to a superconducting instability~\cite{Gurevich_1962,Grabowski_1984,Sham_1985,Pimenov_2022_npj,Pimenov_2022,McMillan_1968,SM}.


One last ingredient is needed in our theory --- dynamical screening~\cite{Pines_and_Nozieres,Mahan_2000,Giuliani_and_Vignale}, which 
captures the fact that electrons in a 2DES move on time scales on the order of $\tau_{\rm p}(q) = 1/\omega_{\rm p}(q)$, where $\hbar\omega_{\rm p}(q)$ is the 2D plasmon dispersion, to screen external electrical disturbances. In weakly correlated 2D materials such as graphene and TBG away from $\theta^\star$, the RPA~\cite{Pines_and_Nozieres,Mahan_2000,Giuliani_and_Vignale} captures dynamical screening very well. In the RPA, the effective, dynamically-screened, electron-electron interaction $W(q,\omega)$ is given by
\begin{equation}
W(q,\omega) = \frac{V(q,\omega)}{1 - V(q,\omega)\chi_0(q,\omega)}~,
\end{equation}
where $\chi_{0}(q, \omega)$ is the non-interacting density-density response function of a MDF 2DES~\cite{Wunsch_NJP_2006,Hwang_PRB_2007,Barlas_PRL_2007,Polini_prb_2008}, which is proportional to the number $N_{\rm f}$ of fermion flavors.

\noindent{\color{blue}\it Superconductivity in the presence of repulsive interactions}.---The superconducting instability~\cite{Pimenov_2022} in our model is complicated by the interplay between its three different energy scales~\cite{SM}: i) the longitudinal phonon energy scale $\hbar \omega_{\rm L}$ entering the problem through $V(q,\omega)$, ii) the Fermi energy $\varepsilon_{\rm F}$ of the 2DES entering the problem through $\chi_0(q,\omega)$, and iii) the 2DES plasmon energy scale $\hbar\omega_{\rm p}(q)$ entering the problem through the sharp pole of $W(q,\omega)$ located  below the real-frequency axis.   
In order to find the superconducting critical temperature $T_{\rm c}$ we resort to Eliashberg theory~\cite{Marsiglio_2020}. Neglecting self-energy corrections~\cite{Pimenov_2022}, which are relevant in MDF 2DESs only near the charge neutrality point~\cite{elias_naturephys_2011}, the linearized gap equation for superconductivity in the $\ell$-th angular-momentum channel is 
\begin{multline}\label{eq:lin-gap-l}
\Delta^{(\ell)}(k,\I\omega_n) =
-k_{\text{B}}T\sum_{n'=-\infty}^{\infty} \int_0^{\infty}\dif k'
\, \frac{k'}{2\PI}
\\
\times
\frac{I^{(\ell)}(k,k',\I\omega_n-\I\omega_{n'})
\Delta^{(\ell)}(k',\I\omega_{n'})}{
(\hbar\omega_{n'})^2 + \xi^2(k')}~,
\end{multline}
where $T$ is temperature,
$\hbar\omega_n = \pi(2n + 1)k_{\text{B}}T$ is a fermionic Matsubara frequency,
$\xi(\bm{k})= \hbar v k -\mu$ is band energy measured from the chemical potential 
$\mu$, and the averaged interaction is defined as
\begin{multline}\label{eq:avg-int-l}
I^{(\ell)}(k,k',\I\Omega)
\equiv \int_0^{2\PI}\frac{\dif\theta}{2\PI}
\cos(\ell\theta)\,
\frac{1 + \cos(\theta)}{2}
\\
\times
W(\sqrt{k^2+k'^2-2kk'\cos\theta\,},\I\Omega)~.
\end{multline}
Note the usual chirality factor $f(\theta) = [1+ \cos(\theta)]/2$ for 2D MDFs.
Henceforth, we will omit the superscript $\ell$,
considering only the $s$-wave, $\ell=0$, case.

Since the gap is an even function of frequency,
i.e.~$\Delta(k,\I\omega)=\Delta(k,-\I\omega)$, we can restrict the Matsubara sum to positive frequencies
\begin{multline}\label{eq:lin-gap-s}
\Delta(k,\I\omega_n)
\\
= -k_{\text{B}}T\sum_{n'=0}^{\infty} \int_0^{\infty}\dif k'
\,
K(k,\I\omega_n,k',\I\omega_{n'})
\Delta(k',\I\omega_{n'})~,
\end{multline}
where the integral kernel is
\begin{multline}\label{eq:lin-ker-s}
K(k,\I\omega_n,k',\I\omega_{n'})
\\
\equiv
\frac{k'}{2\PI}
\,
\frac{ I(k,k',\I\omega_n-\I\omega_{n'})
+ I(k,k',\I\omega_n+\I\omega_{n'}) }{
(\hbar\omega_{n'})^2 + \xi^2(k')}~.
\end{multline}
We numerically solve the $s$-wave linearized gap equation~\eqref{eq:lin-gap-s}
to determine the critical temperature $T_{\text{c}}$. Since Eq.~\eqref{eq:lin-gap-s} is a homogeneous linear integral
equation for $\Delta(k,\I\omega)$, $T_{\rm c}$ is found by searching 
for the temperature at which the largest positive eigenvalue is unity.
A nice account of the physical properties of the superconducting instability in the presence of dynamical repulsive interactions (e.g., the fact that $\Delta$ changes sign as a function of imaginary frequency and related physical consequences) can be found in Ref.~\cite{Pimenov_2022}.
Numerical details will be discussed elsewhere.

\begin{figure}
\centering
\includegraphics{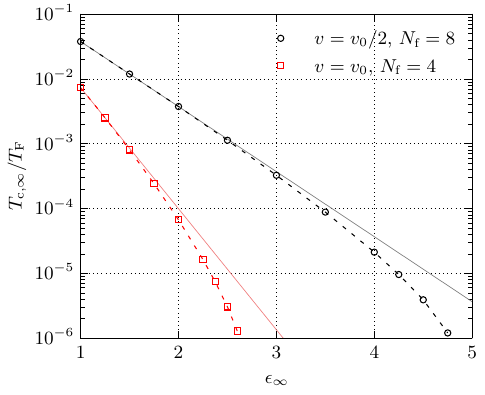}
\caption{Critical temperature $T_{\text{c},\infty}$ (in units of the Fermi temperature $T_{\rm F}$) for plasmon-mediated superconductivity in MDF 2DESs. 
The results in this figure have been obtained by encapsulating the MDF 2DES between two semi-infinite dielectric slabs of permittivity $\epsilon_\infty$. Black circles (red squares) represent numerical results for a MDF 2DES with velocity $v=v_0/2$ ($v=v_0$) and $N_{\text{f}}=8$ ($N_{\text{f}}=4$) flavors, where $v_0=\SI{e6}{m/s}$ is the single-layer graphene Fermi velocity.  ($N_{\text{f}}$ enters the problem via the bare polarization function $\chi_0(q,\omega)$.)
Thin solid lines  illustrate exponential depedence on $\epsilon_\infty$. Our numerical results depart from simple exponential  dependence at weak coupling, i.e. for large $\epsilon_\infty$.}\label{fig:three}
\end{figure}

Eq.~\eqref{eq:lin-gap-l} neglects vertex corrections. Migdal pointed out~\cite{migdal1958interaction} that this approximation can be rigorously 
justified when 
the structure in $I^{(\ell)}(k,k',\I\Omega)$ occurs on a frequency scale that is small compared to the Fermi energy. Recent accurate comparisons~\cite{Chubukov_2020} between the results of Eliashberg theory in the Migdal approximation and quantum Monte Carlo calculations for the 2D Holstein model have pointed out that the smallness of vertex corrections also
requires that the coupling constant of the theory are small. Assuming that the results of Ref.~\cite{Chubukov_2020} have a general character, the smallness of vertex corrections in our problem is therefore controlled by the smallness of the dimensionless coupling constant of the theory, i.e.~$\alpha_\omega \equiv e^2/(\hbar v \epsilon(\omega))$ {\it and} the smallness of the ratio $\hbar\omega_\ast/\varepsilon_{\rm F}$ where $\hbar\omega_\ast$ is the characteristic energy scale of the boson mediating pairing. 

We first solve the Eliashberg equation neglecting the phonon-polariton correction to 
$V(q, \I \Omega)$, i.e.~using $V(q, \I\infty)$ rather than $V(q, \I \Omega)$. In this limit, the superconducting pairing is mediated by the 2D plasmon only, and the corresponding value of the critical temperature will be dubbed $T_{\text{c},\infty}$. In Fig.~\ref{fig:two} we plot (green dotted line) the $s$-wave (i.e.~$\ell=0$) averaged interaction $I(k, k^\prime, \I\Omega)$, defined in Eq.~(\ref{eq:avg-int-l}), corresponding to the instantaneous potential $V(q,\I\infty)$. Since superconductivity is a Fermi surface effect, in making this plot we choose $k = k_{\rm F}$ and $k^\prime = 1.1k_{\rm F}$, where $k_{\rm F}$ is the Fermi wave number. We see that dynamical screening
reduces $I(k, k^\prime, \I\Omega)$ from its high-frequency bare value on the Fermi energy
$\varepsilon_{\rm F}$ scale, even though long-wavelength 2D plasmons have an energy that is much smaller than $\varepsilon_{\rm F}$. Results of the vertex-free theory therefore should not be trusted for values
of $\alpha_\infty = e^2/(\hbar v \epsilon_\infty) \approx 2.2 v_0/(v \epsilon_\infty)$
that are ${\cal O}(1)$,
and are also uncertain at smaller values of $\alpha_\infty$ even though 
$\hbar \omega_\ast/\varepsilon_{\rm F} = \hbar\omega_{\rm p}(q_\ast)/\varepsilon_{\rm F}<1$ at some $q_\ast < k_{\rm F}$.

We propose that limits on the vertex-correction-free plasmon-mediated superconductivity theory be established by comparing the predictions summarized in Fig.~\ref{fig:three} with experiment. In Fig.~\ref{fig:three} we have plotted the ratio of $T_{{\rm c}, \infty}$ to the 
Fermi temperature $T_{\rm F}$ as a function of the high-frequency dielectric permittivity $\varepsilon_{\infty}$ for a MDF 2DES encapsulated between two semi-infinite dielectric slabs,
($d_{\text{b}}=d_{\text{t}}\to\infty$)
of permittivity $\epsilon_\infty$.
Consider first the case of single-layer graphene ($N_{\text{f}}=4$ and $v=v_0=\SI{e6}{m/s}$). When this is encapsulated in a material with $\epsilon_\infty=4$,
we do not find a numerical solution of the gap equation,
consistent with the experimental absence of superconductivity in the case of single-layer graphene encapsulated in e.g.~hBN. Suspended single-layer graphene, i.e.~$\epsilon_\infty=1$,
corresponds to the leftmost point in Fig.~\ref{fig:three}.
In this case, we find that $T_{\text{c},\infty}/T_{\text{F}}\sim\num{7e-3}$,
not in obvious contradiction with experiments since the Fermi temperatures
that are achievable in suspended samples are relatively small~\cite{High_Density_Suspended_Graphene}. Our predictions for $T_{{\rm c}, \infty}/T_{\rm F}$ in the case of TBG ($N_{\text{f}}=8$ and $v=v_0/2=0.5 \times \SI{e6}{m/s}$, corresponding to a twist angle $\theta \simeq 2.2^\circ$) are also reported in Fig.~\ref{fig:three}. The vertex-free theory could be tested quantitatively by preparing relatively high density TBG encapsulated devices in which the velocity is reduced, increasing $\alpha_{\infty}$. If the vertex-free theory is accurate, superconductivity should then be readily observable. We now proceed to study the role of the polaritonic medium surrounding the MDF 2DES. As we will argue below, we believe that the vertex-free theory, which predicts 
enhanced transition temperatures, is under better control in this case.

\begin{figure}[t]
\centering
\begin{overpic}[percent]{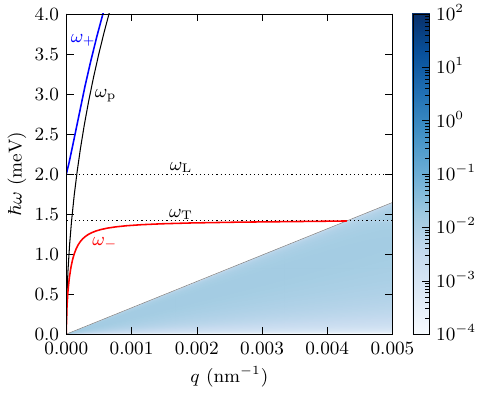}
\put(0,80){\makebox(0,0)[tl]{(a)}}
\end{overpic}
\begin{overpic}[percent]{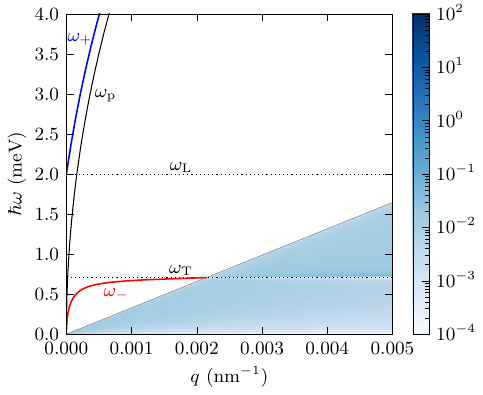}
\put(0,80){\makebox(0,0)[tl]{(b)}}
\end{overpic}
\caption{
Plasmon-phonon polaritons. The spectral function, 
$-\Im W(q,\omega)$, of the dynamically-screened electron-electron interaction (in units of $2\pi e^2/k_{\text{F}}$) is plotted as a function of wave vector $q$ and real frequency $\omega$.
The blue and red solid lines represent the upper and lower polaritons 
$\omega_{\pm}(q)$ outside the intra-band electron-hole continuum of the 2DES.
The solid black line represents the dispersion of the bare plasmon
$\omega_{\text{p}}(q)$ of the 2DES.
The dotted black lines represent the longitudinal and transverse optical
phonon modes of the dielectric slabs ($\omega_{\text{L}}>\omega_{\text{T}}$).
In this plot the energy of the longitudinal optical phonon mode has been set at
$\hbar\omega_{\text{L}}=\SI{2}{meV}$. All the other parameters are as in Fig.~\ref{fig:one}.
Panel (a) refers to $\epsilon_0=6$ ($\zeta=\sqrt{2}/2$)---see data labelled by red circles in Fig.~\ref{fig:one}. 
Panel (b) to $\epsilon_0=24$ ($\zeta=\sqrt{7/8}$)---see data labelled by blue squares in Fig.~\ref{fig:one}.}\label{fig:four}
\end{figure}

\noindent{\color{blue}\it Polaritonic enhancement of $T_{\rm c}$}.---In the main panel of Fig.~\ref{fig:one} we report the ratio $T_{\rm c}/T_{\rm F}$ for a MDF 2DES with $N_{\rm f}=8$, velocity $v = v_0/2 \simeq 0.5 \times 10^{6}~{\rm m}/{\rm s}$, and carrier density $n = 10^{11}~{\rm cm}^{-2}$. In this case, $T_{\rm F} \approx 150~{\rm K}$. (In the TBG platform, this value of $v/v_0$ corresponds to a twist angle $\theta \simeq 2.2^\circ$.) As in the case of Fig.~\ref{fig:three}, we assume that the MDF 2DES is encapsulated between two semi-infinite dielectric slabs (i.e.~$d_{\rm t} = d_{\rm b} \to \infty$). 
In Fig.~\ref{fig:one} $T_{\rm c}/T_{\rm F}$ is plotted as a function of the longitudinal phonon energy scale $\hbar\omega_{\rm L}$ at two 
different values of the phonon oscillator strength $\zeta$.  The left  
limit of these curves corresponds to 2D plasmonic superconductivity
at dielectric constant $\epsilon_{\infty}=3$, whereas the right limit corresponds
to 2D plasmonic superconductivity at dielectric constants $\epsilon_{0}=6, 24$. The maximum in the 
predicted $T_{\rm c}$ occurs when $\hbar \omega_{\rm L} \sim 0.15 \, \varepsilon_{\rm F}$.
We note three relevant facts: a) Because of dielectric screening, $T_{\rm c}$
is reduced with respect to $T_{{\rm c}, \infty}$ when $\hbar \omega_{\rm L} > \varepsilon_{\rm F}$; b) There is nevertheless 
a clear range of values of $\hbar \omega_{\rm L}$ near the curve maxima where 
$T_{\rm c}$ is increased by coupling to the dielectric; c) Finally, the larger the phonon oscillator strength $\zeta$, the more pronounced the enhancement of $T_{\rm c}$. In Fig.~\ref{fig:one} we see an enhancement of $T_{\rm c}$ on the order of $40\%$ for $\zeta = \sqrt{7/8}$ relative to the plasmonic $T_{\rm c}$ calculated with $\epsilon(\omega) \to \epsilon_{\infty}$, i.e.~relative to $T_{{\rm c}, \infty}$. 

A natural question arises: What is the reason for the enhancement of $T_{\rm c}$ with respect to the bare value $T_{{\rm c}, \infty}$? The answer is that it is entirely due to the resonant interaction between the two bare modes of the heterostructure, i.e.~the longitudinal optical phonon in the polar material occurring at an energy $\hbar \omega_{\rm L}$ and the plasmon mode of the MDF 2DES at energy $\hbar \omega_{\rm p}(q)$. When these two energy scales are resonant, two new modes emerge, an upper $\hbar\omega_+(q)$ and a lower $\hbar\omega_-(q)$ plasmon-phonon polariton. This is shown in Fig.~\ref{fig:four}. Comparing panel (a) with panel (b), we clearly see that the larger the oscillator strength of the longitudinal optical phonon, the bigger is the avoided crossing (i.e.~hybridization gap) between the two modes. This means that the lower polariton $\hbar\omega_-(q)$, which is a quasi-bosonic excitation, shifts spectral weight from high energies to low energies, yielding an enhancement of $T_{\rm c}$ with respect to $T_{{\rm c}, \infty}$.  As we explain below, the enhancement is due to structure introduced in the retarded interaction at frequencies below the 
Fermi energy which is similar to the structure introduced by intrinsic phonons in typical bulk superconductors and therefore reliably described by Eliashberg theory.

In Fig.~\ref{fig:two} we plot averaged interaction $I(k_{\rm F},1.1 k_{\rm F}, \I \Omega)$ in the $s$-wave channel as a function of the imaginary frequency $\Omega$, for the case in which polaritons are present. Three curves are reported in this figure, for different values of the longitudinal phonon energy $\hbar \omega_{\rm L}$. We start by describing the case in which $\hbar \omega_{\rm L} = 200~{\rm meV}$. In this case the longitudinal optical phonon exceeds the Fermi energy $\varepsilon_{\rm F}$ and the 2DES plasmon energy $\hbar\omega_{\rm p}(q)$. We clearly see two smooth ``steps'' in $I(k_{\text{F}},1.1 k_{\text{F}},\I \Omega)$, one at the $\varepsilon_{\rm F}$ scale and another step at $\hbar\omega_{\rm L}$. The later reduction reduces the 
effective dielectric constant for plasmonic superconductivity from 
$\epsilon_{\infty}$ to $\epsilon_0$, leading to a dramatic suppression of $T_{\rm c}$ with respect to $T_{{\rm c}, \infty}$ (see Fig.~\ref{fig:one}), but the superconductivity is still plasmonic in character. For $\hbar \omega_{\rm L} = 0.2~{\rm meV}$ on the other hand, the phonon energy scale is negligible with respect to the $\varepsilon_{\rm F}$ scale. A small shoulder appears in $I(k_{\text{F}},1.1 k_{\text{F}},\I \Omega)$ at $\Omega=\omega_{\rm L}$ 
but it has no impact on the plasmonic superconductivity.  This is the regime in
which $T_{\rm c}$ is close to the value calculated without the polaritonic coupling
but using dielectric constant $\epsilon_{\infty}$ (see Fig.~\ref{fig:one}). 
Finally, we analyze the case in which $\hbar \omega_{\rm L} = 2~{\rm meV}$. In this case the longitudinal optical phonon of the dielectric slabs is strongly resonant with the 2DES plasmon over a substantial range of $q$---see Fig.~\ref{fig:four}. We clearly see that, in this case, $I(k_{\text{F}},1.1 k_{\text{F}},\I \Omega)$ has a single smooth ``step'' as a function of frequency, which optimizes the window of McMillan attraction. At this value of $\hbar \omega_{\rm L}$, $T_{\rm c}$ is maximized---see Fig.~\ref{fig:one}---and 
vertex corrections are expected to be weak. Indeed, we note two key facts: i) dielectric screening due to the polaritonic medium reduces the value of the dimensionless coupling constant with respect to the case of suspended samples and ii) in the case of a resonant coupling between the 2DES plasmon and polaritons, the characteristic energy scale $\hbar\omega_\ast$ is suppressed with respect to the case in which polaritons are not present since $\hbar\omega_\ast = \hbar\omega_-(q_\ast)\ll \hbar \omega_{\rm p}(q_\ast)$ at the same $q_\ast$.
We predict that 
observable engineered superconductivity will occur in this regime.

In this Letter we have presented a minimal model whereby polaritons aid superconductivity, yielding a significant increase in the superconducting critical temperature. (The role of hyperbolicity is analyzed in Ref.~\cite{SM}.)  Experimental tests of this prediction require placing MDFs in the environment of polar dielectrics with longitudinal
optical phonons that have large oscillator strengths and energies below the 
MDF Fermi energy. We believe it is highly desirable to 
fabricate twisted graphene bilayers with reduced Dirac velocities.  The resulting 
increase in the Fermi level density-of-states plays a key role in yielding critical temperatures that are significantly larger than in single-layer graphene. As the typical Fermi energies are reduced, however, the list of 
qualified dielectrics shrinks to those with phonon polaritons in the THz spectral range. Detailed calculations for the case of specific materials will be published elsewhere.
A further boost in $T_{\rm c}$ may come from creating a resonant structure where the dielectric slabs encapsulating the MDF 2DES are spatially patterned to create a polaritonic lattice~\cite{sheinfux_naturematerials_2024}. Indeed, spatial patterning can create a resonant interaction not only in the $\omega$ domain, as we have seen above, but also in momentum space. Tuning the periodicity of the lattice one may be able to create mode matching at selected hot spots in the $\omega$-${\bm q}$ plane where the spectral density of states is largest.

{\color{blue} {\it Acknowledgments}.}---We thank Gian Marcello Andolina for useful discussions.
R.R. was supported by the
``National Centre for HPC, Big Data and Quantum Computing''
under the National Recovery and Resilience Plan (NRRP),
Mission 4 Component 2 Investment 1.4 CUP I53C22000690001
funded from the European Union -- NextGenerationEU.

F.H.L.K., and M.P. were supported by the European Union under grant agreement No. 101131579 - Exqiral. 

F.H.L.K. acknowledges support from the Gordon and Betty Moore Foundation through Grant
  GBMF12212, and the Government of Spain (QTWIST RD0768/2022, PID2022-141081NB-I00; Severo Ochoa
  CEX2019-000910-S, and CEX2024-001490-S [MCIN/AEI/10.13039/501100011033]). This work was also
  supported by the European Union NextGenerationEU/PRTR (PRTR-C17.I1),
  Fundació Cellex, Fundació Mir-Puig, Generalitat de Catalunya (CERCA, Department of Digital
  Policies and Territory, AGAUR, 2021 SGR 014431656).  This material is based upon work supported by the Air Force Office of Scientific
  Research under Award Number FA8655-23-1-7047. 

P.J.H. acknowledges support by AFOSR grant FA9550-21-1-0319, ONR grant N000142412440, the MIT/Microsystems Technology Laboratories Samsung Semiconductor Research Fund, the Gordon and Betty Moore Foundations EPiQS Initiative through Grant GBMF9643, the Ramon Areces Foundation, and the CIFAR Quantum Materials program. 

G.M.~acknowledges support by the MUR Italian 
Minister of University and Research through a ``Rita Levi-Montalcini'' fellowship. 

A.H.M. was supported by the W. M. Keck Foundation under grant 996588.

Views and opinions expressed in this material are those of the author(s) only and do not necessarily reflect those of the European Union or the European Commission. Neither the European Union nor the granting authority can be held responsible for them.

 Any opinions, findings, conclusions, or
  recommendations expressed in this material are those of the author(s) and do not necessarily
  reflect the views of the United States Air Force.

\appendix
\clearpage 
\setcounter{section}{0}
\setcounter{equation}{0}%
\setcounter{figure}{0}%
\setcounter{table}{0}%

\setcounter{page}{1}

\renewcommand{\thetable}{S\arabic{table}}
\renewcommand{\theequation}{S\arabic{equation}}
\renewcommand{\thefigure}{S\arabic{figure}}
\renewcommand{\bibnumfmt}[1]{[S#1]}

\onecolumngrid

\begin{center}
\textbf{\Large Supplemental Material for
``Polariton-induced superconductivity in two-dimensional metals''}
\bigskip

Riccardo Riolo,${}^{1}$
Frank H.L. Koppens,${}^{2,3}$
Pablo Jarillo-Herrero,${}^{4}$
Giacomo Mazza,${}^{1}$
Allan H. MacDonald,${}^{5}$
and Marco Polini${}^{1,2}$

\bigskip

${}^1$\textit{Dipartimento di Fisica dell’Università di Pisa, Largo Bruno Pontecorvo 3, I-56127 Pisa, Italy}

${}^2$\textit{ICFO-Institut de Ciències Fotòniques, The Barcelona Institute of Science and Technology,\\
Av. Carl Friedrich Gauss 3, 08860 Castelldefels (Barcelona), Spain}

${}^3$\textit{ICREA-Institució Catalana de Recerca i Estudis Avançats, Barcelona, Spain}

${}^4$\textit{Department of Physics, Massachusetts Institute of Technology, Cambridge, Massachusetts, USA}

${}^5$\textit{Department of Physics, The University of Texas at Austin, Austin, TX 78712, USA}

\bigskip

In this Supplemental Material we provide additional information on an analytical treatment of the gap equation and additional numerical results on the role of metal gates and hyperbolicity of the dielectrics.
\end{center}

\twocolumngrid

\section{Quick recapitulation of the McMillan formula for the critical temperature of strongly-coupled superconductors}

Our analytical approach to solve Eq.~(\ref{eq:lin-gap-s}) in the main text is inspired by the derivation of the McMillan formula for the critical temperature of strongly-coupled superconductors, which we here quickly recapitulate.

For strongly-coupled \emph{phonon} superconductors, the Eliashberg equations read as following~\cite{Marsiglio_2020}:
\begin{widetext}
\begin{gather}
Z(\bm{k},\I\omega_n)
= 1 - \frac{k_{\text{B}}T}{V} \sum_{n',\bm{k}'}
W(\bm{k} - \bm{k}',\I\omega_n - \I\omega_{n'})
\,\frac{(\omega_{n'}/\omega_{n})
Z(\bm{k}',\I\omega_{n'})}{
(\hbar\omega_{n'}Z(\bm{k}',\I\omega_{n'}))^2
+ \xi^2(\bm{k}')
+ \Phi^2(\bm{k}',\I\omega_{n'})}~,
\label{eq:Eli_1}\\
\Phi(\bm{k},\I\omega_n)
= - \frac{k_{\text{B}}T}{V} \sum_{n',\bm{k}'}
W(\bm{k} - \bm{k}',\I\omega_n - \I\omega_{n'})
\,\frac{\Phi(\bm{k}',\I\omega_{n'})}{
(\hbar\omega_{n'}Z(\bm{k}',\I\omega_{n'}))^2
+ \xi^2(\bm{k}')
+ \Phi^2(\bm{k}',\I\omega_{n'})}~,\label{eq:Eli_2}
\end{gather}
\end{widetext}
where
$Z(\bm{k},\I\omega)$ is defined from the normal self-energy $\Sigma(\bm{k},\I\omega)$ as
\begin{equation}
\I\hbar\omega[1 - Z(\bm{k},\I\omega)]
\equiv \frac{\Sigma(\bm{k},\I\omega) - \Sigma(\bm{k},-\I\omega)}{2},
\end{equation}
$\Phi(\bm{k},\I\omega)$ is the anomalous self-energy,
and $W(\bm{k},\I\omega)$ is the effective electron-electron interaction.
The equation for the gap
$\Delta(\bm{k},\I\omega)=\Phi(\bm{k},\I\omega)/Z(\bm{k},\I\omega)$
is usually simplified~\cite{Marsiglio_2020}
by neglecting the wave-vector dependence
of the gap function, $\Delta(\bm{k}, \I\omega) \to \Delta(\I \omega)$, and of the electron-electron interaction $W(\bm{k}, \I \omega) \to W(\I\omega)$.
By replacing the integration over the wave vector $\bm{k}'$ in Eqs.~(\ref{eq:Eli_1})-(\ref{eq:Eli_2}) with an integration over the energy $\xi$,
and approximating the density of states as the density of states at the Fermi surface,
$\nu(\xi)\approx\nu(0)\equiv\nu_{\text{F}}$,
the Eliashberg equations reduce to~\cite{Marsiglio_2020}
\begin{gather}
Z(\I\omega_n)
= 1 - \frac{\pi k_{\text{B}}T}{\hbar\omega_n} \sum_{n'}
\frac{\nu_{\text{F}}W(\I\omega_n - \I\omega_{n'}) \hbar\omega_{n'}}{
\sqrt{(\hbar\omega_{n'})^2 + \Delta^2(\I\omega_{n'})}}~.\label{eq:Eli_3}
\\
\Delta(\I\omega_n)
= -\frac{\pi k_{\text{B}}T}{Z(\I\omega_n)} \sum_{n'}
\frac{\nu_{\text{F}}W(\I\omega_n - \I\omega_{n'}) \Delta(\I\omega_{n'})}{
\sqrt{(\hbar\omega_{n'})^2 + \Delta^2(\I\omega_{n'})}}~.\label{eq:Eli_4}
\end{gather}
The electron-electron interaction $W$ includes an attractive contribution
due to the electron-phonon interaction and a repulsive contribution due
to the Coulomb interaction.
Considering a single Einstein phonon mode of energy $\hbar\omega_0$ and a Hubbard-like local and instantaneous repulsion $U$,
the interaction is
\begin{equation}\label{eq:hubbard-einstein}
W(\I\Omega) = -g^2\,\frac{2\hbar\omega_0}{(\hbar\Omega)^2 + (\hbar\omega_0)^2} + U~,
\end{equation}
where $g$ is the electron-phonon coupling (with units of energy times the square root of a volume).

For strong electron-phonon coupling, we need to consider the normal self-energy part due to phonons.
From the simplified Eliashberg equation for $Z(\I\omega_n)$, i.e.
\begin{multline}
Z(\I\omega_{n})
= 1 + \frac{\pi k_{\text{B}}T}{\hbar\omega_{n}}\sum_{n'}
\frac{2\nu_{\text{F}}g^2\hbar\omega_0}{(\hbar\omega_n-\hbar\omega_{n'})^2 + (\hbar\omega_0)^2} \\
\times
\frac{\hbar\omega_{n'}}{
\sqrt{(\hbar\omega_{n'})^2 + \Delta^2(\I\omega_{n'})}}~,
\end{multline}
one finds~\cite{McMillan_1968}, after linearization and upon carrying out straightforward algebraic manipulations, that  the low- and high-frequency values of $Z$ are $Z_0 = 1 + \lambda$ and $Z_\infty = 1$, respectively.

An approximate analytical solution of the linearized gap equation can be found~\cite{McMillan_1968} by
chosing a trial function for the gap $\Delta(\I\omega)$.
The gap function is approximated by a piecewise constant function~\cite{McMillan_1968}:
\begin{equation}
\Delta(\I\omega) =
\begin{cases}
\Delta_0, & 0 < \lvert \omega \rvert < \omega_0~, \\
\Delta_\infty, & \omega_0 < \lvert \omega \rvert~.
\end{cases}
\end{equation}
As a further simplification, the same piecewise-constant approximation can be applied to the interaction~\cite{Rietschel_1983}:
\begin{equation}
\nu_{\text{F}}W(\I\omega) =
\begin{cases}
-\lambda + u, & 0 < \lvert \omega \rvert < \omega_0~, \\
u, & \omega_0 < \lvert \omega \rvert~,
\end{cases}
\end{equation}
where $\lambda = 2\nu_{\text{F}}g^2/\hbar\omega_0$ is the electron-phonon
coupling constant
and $u = \nu_{\text{F}} U$ is the Hubbard repulsive coupling.

Replacing the trial function into the right-hand side of the gap
equation (\ref{eq:Eli_4}) and performing the Matsubara sum,
we impose that the resulting left-hand
side is consistent with the chosen trial function.
As noted by Morel and Anderson~\cite{Morel_1962},
the substitution of the integration over the wave vector $\bm{k}'$ with an integration over $\xi$ in the simplified gap equation
introduces an artificial divergence in the Matsubara sum
in the case of a local and instantaneous interaction $U$.
This divergence can be cured by introducing a cut off
$\Lambda$ on the order of the Fermi energy, i.e.~$\Lambda\sim\varepsilon_{\text{F}}$.
The linearized gap equation can be thus recasted into the following linear system of equations~\cite{McMillan_1968}:
\begin{gather}
(1 + \lambda) \Delta_0
= (\lambda - u) \Delta_0 \ln\frac{\hbar\omega_0}{k_{\text{B}}T_{\text{c}}}
- u \Delta_\infty \ln\frac{\Lambda}{\hbar\omega_0}~,
\\
\Delta_\infty
= - u \Delta_0 \ln\frac{\hbar\omega_0}{k_{\text{B}}T_{\text{c}}}
- u \Delta_\infty \ln\frac{\Lambda}{\hbar\omega_0}~.
\end{gather}

Since the system has a non-zero solution at the critical temperature, its determinant must vanish. This yields the following equation for $T_{\text{c}}$:
\begin{equation}
k_{\text{B}} T_{\text{c}} \sim \hbar \omega_0
\exp\biggl(-\frac{1 + \lambda}{\lambda - u^*}\biggr)~,
\end{equation}
where $u^*$ is the Morel--Anderson pseudopotential~\cite{Morel_1962}
\begin{equation}
u^* \equiv \frac{u}{
1 + u \ln(\Lambda / \hbar \omega_0)}~.
\end{equation}
This shows that the repulsive potential is renormalized because of the
frequency dependence of the interactions and it is
reduced, $u^*<u$, i.e.~it is less effective.

An analogous pairing mechanism is also at play in systems with
dynamical repulsive interactions~\cite{Gurevich_1962,Pimenov_2022_npj,Pimenov_2022}.
Approximating the interaction as~\cite{Rietschel_1983}
\begin{equation}
\nu_{\text{F}}W(\I\omega) =
\begin{cases}
u_0, & 0 < \lvert \omega \rvert < \omega_0, \\
u_\infty, & \omega_0 < \lvert \omega \rvert,
\end{cases}
\end{equation}
and neglecting the normal self-energy corrections,
i.e.~setting $Z(\I\omega)\equiv 1$,
the gap equation can be recast as
\begin{gather}
\Delta_0
= - u_0 \Delta_0 \ln\frac{\hbar\omega_0}{k_{\text{B}}T_{\text{c}}}
- u_\infty \Delta_\infty \ln\frac{\Lambda}{\hbar\omega_0},
\\
\Delta_\infty
= - u_\infty \Delta_0 \ln\frac{\hbar\omega_0}{k_{\text{B}}T_{\text{c}}}
- u_\infty \Delta_\infty \ln\frac{\Lambda}{\hbar\omega_0}.
\end{gather}
A non-zero solution is possible only if
\begin{equation}
u_0<\frac{u_\infty^2\ln(\Lambda/\hbar\omega_0)}{1 + u_\infty\ln(\Lambda/\hbar\omega_0)}<u_\infty,
\end{equation}
thus, a necessary condition is that the interaction is less repulsive at lower frequencies
than at higher frequencies.
In this case, the critical temperature is
\begin{equation}
k_{\text{B}}T_{\text{c}} \sim \hbar\omega_0
\exp\biggl(-\frac{1}{(u_\infty - u_0)
- \frac{u_\infty}{1 + u_\infty \ln(\Lambda/\hbar\omega_0)}}\biggr).
\end{equation}
Note that, in this case, the difference $u_\infty - u_0$ acts as an effective attractive
coupling (this is what we dubbed in the main text ``window of McMillan attraction'') and the high-frequency repulsive interaction $u_\infty$ is
renormalized as in the case of the Morel--Anderson pseudopotential. (The condition $u_0<u_\infty$ is of course satisfied by the
electron-phonon interaction with an instantaneous Hubbard-$U$ repulsive
interaction discussed above. 
The same window $u_\infty - u_0$ of attraction applies in the case of a purely
repulsive, dynamical interaction, which is less repulsive at lower frequencies than at higher frequencies.)

In Ref.~\cite{Pimenov_2022},
the interaction in Eq.~\eqref{eq:hubbard-einstein} (there called HEF) is also considered without employing the piecewise-constant approximation.
The interaction can be parametrized as
\begin{equation}\label{eq:HEF}
\nu_{\text{F}} W_{\text{HEF}}(\I\omega)
= \lambda \biggl( f
- \frac{\omega_0^2}{\omega + \omega_0^2} \biggr),
\end{equation}
where $\omega_0$ is the frequency of the Einstein phonon, $\lambda$ is the coupling strength
and $f$ is the Hubbard-like repulsion strength.
For $\omega\ll\omega_0$, the interaction strength is $\lambda f - \lambda$,
while for $\omega\gg\omega_0$, the interaction strength is $\lambda f$.
For $f<0$, the interaction is always attractive on the imaginary axis of the frequencies, and the gap function is nodeless.
For $0<f<1$, the interaction is attractive for small $\omega$ but is repulsive for larger $\omega$.
In this case, the gap function presents a node on the imaginary axis of the frequencies.
For $f\ge1$, the interaction is purely repulsive and superconductivity develops only if the coupling $\lambda$ exceeds a critical $\lambda_{\text{c}}$, where
\begin{equation}
\lambda_{\text{c}}
= \frac{f - 1}{2f\ln(\Lambda/\hbar\omega_0)}.
\end{equation}
Inverting this relation for fixed $\lambda$,
superconductivity develops for $f$ below a critical repulsion strength, $f<f_{\text{c}}$, where
\begin{equation}
f_{\text{c}}
= \frac{1}{1 - 2\lambda\ln(\Lambda/\hbar\omega_0)}.
\end{equation}
For weak coupling, the superconducting temperature near criticality is exponentially small,
$T_{\text{c}}\sim \exp(-1/(f - f_{\text{c}}))$.

In the main text, we consider the plasmon mechanism of superconductivity, where the screened interaction $W$ presents a frequency dependence analogous to the HEF model (see Fig.~\ref{fig:two} of the main text).
The strength of the interaction $\lambda$ is proportional to the electron-electron coupling constant $\alpha_\infty = e^2/\hbar v\epsilon_\infty$.
Because of this frequency structure of the interaction, we expect that, in the weak-coupling regime, the superconducting transition cannot occur below a threshold $\alpha_\infty^*$.
However, the parameter $f$ is determined by the difference between the high- and low-frequency values of the interaction, which are dependent both on the coupling $\alpha_\infty$ and on the exchanged wave vector (since we consider the long-range Coulomb interaction).
The same applies to the characteristic energy $\hbar\omega_0$, corresponding to the plasmon energy, which depends both on the coupling and the wave vector.
In our numerical solution of the gap equation, we fully consider the wave vector dependence of the gap function, thus we cannot make a one-to-one correspondence with the HEF model parametrization.

On the other hand, Grabowski and Sham~\cite{Grabowski_1984} first
proposed a simplification of the gap equation that leads to the HEF model.
They simplified the interaction by averaging over wave vectors, thus obtaining an interaction that depends only on the frequency, with the functional form of Eq.~\eqref{eq:HEF}.
In this approach, the gap equation is simplified but the parameters $\lambda$, $f$, and $\omega_0$ depend on the microscopic details of the system and are found by averaging the interaction over wave vectors.
Grabowski and Sham considered the plasmon mechanisms for the 3D electron gas, where there is a gapped optical plasmon. They found that the critical temperature in this approximation is in agreement with the numerical solution of the gap with the wave vector dependence.

In the plasmon mechanism, the dependence of the critical temperature on the coupling $\alpha_\infty$ is not trivial.
By analogy with the purely repulsive HEF model, we expect that also the system of 2D massless Dirac fermions has a critical coupling below which there is no superconductivity.
In the numerical search for $T_{\text{c}}$, we did not find a solution for small coupling.
However, it is difficult to precisely extract this threshold from the numerical data, due to the exponentially small values of $T_{\text{c}}/T_{\text{F}}$ near criticality.
The precise determination of the scaling of the critical temperature with the coupling is beyond the scope of this work.

Remark: a naive formula for the critical temperature $T_{\text{c}}=c_1 T_{\text{F}} \exp(-c_2/(\alpha_\infty - \alpha_\infty^*))$ does not fit the numerical data well for $T_{\text{c}}$. As stated above, this is due to the fact that the repulsion strength $f$ depends on $\alpha_\infty$ and the exchanged wave vector.

\section{Approximate analytic solution of the gap equation for polaritonic-assisted superconductivity}

To find an approximate solution of the linearized gap equation~\eqref{eq:lin-gap-s} in the main text,
we follow the procedure of McMillan~\cite{McMillan_1968},
using a trial function for the gap $\Delta(k,\I\omega)$. 
In our case, pairing arises from the dynamical nature of repulsive
electron-electron
interactions~\cite{Grabowski_1984,Sham_1985}.
We thus consider the frequency dependence of the gap function in Eq.~\eqref{eq:lin-gap-s}.
This pairing is allowed by the less repulsive nature of the
interaction at lower frequencies~\cite{Pimenov_2022}.
For the sake of simplicity,
we assume that the 2DES of MDFs is encapsulated between two
{\it semi-infinite} polar dielectric media, whose permittivity is
given in Eq.~\eqref{eq:simple_permittivity}.

For this simplified geometry, the screened electron-electron interaction on the imaginary-frequency axis is
\begin{equation}
W(q,\I\omega) = \biggl[
\frac{q\epsilon(\I\omega)}{2\PI e^2}
- \chi^{(0)}(q,\I\omega)
\biggr]^{-1}~.
\end{equation}
Such screened interaction is less repulsive at lower frequencies and
increases in a significant fashion at {\it two} energy scales, i.e.~at $\hbar\omega_{\text{L}}$ and $\varepsilon_{\text{F}}$,
respectively corresponding to the longitudinal optical phonon energy of the dielectrics and the Fermi energy of the 2DES (the latter controlling
the low-energy plasmon mode of the 2DES).

\begin{figure}
\centering
\includegraphics{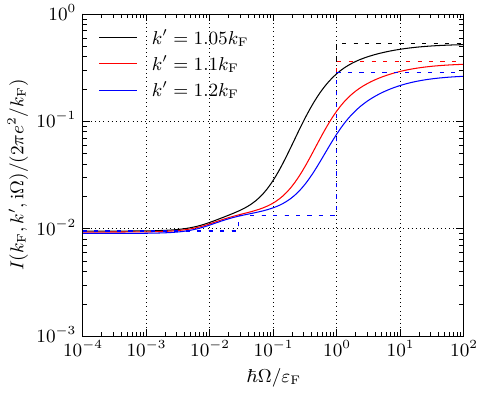}
\caption{%
Averaged interaction $I(k_{\text{F}},k',\I\Omega)$ as a function of the frequency $\Omega$, for different values of $k'$.
Solid lines represent the average interaction in Eq.~\eqref{eq:avg-int-l}, while dashed lines represent the piecewise constant approximation.
The system is sketched in Fig.~\ref{fig:one}.
The dielectric has $\epsilon_0=24$, $\epsilon_\infty=3$, and the frequency of the transverse optical phonon mode is $\hbar\omega_{\text{T}}=10^{-2}\varepsilon_{\text{F}}$.
The parameters are:
$d_{\text{b}}=d_{\text{t}}\to\infty$,
$n = \SI{e11}{cm^{-2}}$,
$v=v_0/2$ ($\theta=2.2^\circ$).
}\label{fig:Iomegaapprox}
\end{figure}

We therefore adopt a step-like approximation for the
interaction~\cite{Rietschel_1983},
which, for small wave vectors and
$\hbar\omega_{\text{L}}<\varepsilon_{\text{F}}$, is given by
\begin{equation}
(W(q,\I\Omega))^{-1} =
\begin{dcases}
\frac{\epsilon_0}{2\PI e^2}q + \nu_{\text{F}}~,
& 0<\Omega<\omega_{\text{L}}, \\
\frac{\epsilon_\infty}{2\PI e^2}q + \nu_{\text{F}}~,
& \omega_{\text{L}}<\Omega<\varepsilon_{\text{F}}/\hbar~, \\
\frac{\epsilon_\infty}{2\PI e^2}q,
& \Omega>\varepsilon_{\text{F}}/\hbar~,
\end{dcases}
\end{equation}
while, for $\hbar\omega_{\text{L}}>\varepsilon_{\text{F}}$, is
\begin{equation}
(W(q,\I\Omega))^{-1} =
\begin{dcases}
\frac{\epsilon_0}{2\PI e^2}q + \nu_{\text{F}}~,
& 0<\Omega<\varepsilon_{\text{F}}/\hbar~, \\
\frac{\epsilon_0}{2\PI e^2}q,
& \varepsilon_{\text{F}}/\hbar<\Omega<\omega_{\text{L}}, \\
\frac{\epsilon_\infty}{2\PI e^2}q,
& \Omega>\omega_{\text{L}}~.
\end{dcases}
\end{equation}
Here, $\nu_{\text{F}}=N_{\text{f}}k_{\text{F}}/(2\pi\hbar v)$ is the
density of states at the Fermi energy of a 2D system of MDFs with velocity $v$. In Fig.~\ref{fig:Iomegaapprox}, we report the averaged interaction $I(k,k',\I\Omega)$ as a function of the frequency $\Omega$, and compare it with our drastically-simplified piecewise-constant approximation.

Following McMillan~\cite{McMillan_1968}, we also adopt a piecewise-constant function
as a trial function for the gap.
Substituting this function into the right-hand side of
Eq.~\eqref{eq:lin-gap-s} and performing the integral,
we impose that the resulting left-hand side is consistent with the trial function.
Since the gap $\Delta(k,\I\omega)$ is peaked around the Fermi wave number, we assume that it
has support on $0<k<2k_{\text{F}}$. Assuming $k_{\text{B}}T\ll\hbar\omega_{\text{L}}\ll\varepsilon_{\text{F}}$,
we approximate the gap as
\begin{equation}
\Delta(\I\omega) =
\begin{cases}
\Delta_1~, & \lvert\hbar\omega\rvert < \hbar\omega_{\text{L}}, \\
\Delta_2~, & \hbar\omega_{\text{L}} < \lvert\hbar\omega\rvert < \varepsilon_{\text{F}}, \\
\Delta_3~, & \varepsilon_{\text{F}} < \lvert\hbar\omega\rvert.
\end{cases}
\end{equation}
Performing the integrals in $k'$ and the sums in $\omega'$ in the gap
equation~\eqref{eq:lin-gap-s} with piecewise-constant approximations for the
kernel,
we find the following linear system
\begin{subequations}
\begin{gather}
\begin{multlined}[b]
\Delta_1
= - \alpha_{1,0}\ln\frac{\hbar\omega_{\text{L}}}{k_{\text{B}}T} \Delta_1
\\
- \alpha_{1,\infty}\ln\frac{\varepsilon_{\text{F}}}{\hbar\omega_{\text{L}}} \Delta_2
- \alpha_{3,\infty} \Delta_3~,
\end{multlined}
\\
\begin{multlined}[b]
\Delta_2
= - \alpha_{1,\infty}\ln\frac{\hbar\omega_{\text{L}}}{k_{\text{B}}T} \Delta_1
\\
- \alpha_{1,\infty}\ln\frac{\varepsilon_{\text{F}}}{\hbar\omega_{\text{L}}} \Delta_2
- \alpha_{3,\infty} \Delta_3~,
\end{multlined}
\\
\begin{multlined}[b]
\Delta_3
= \alpha_{2,\infty}\biggl(
\ln^2\frac{\hbar\omega_{\text{L}}}{\varepsilon_{\text{F}}}
- \ln^2\frac{k_{\text{B}}T}{\varepsilon_{\text{F}}} \biggr) \Delta_1 \\
- \alpha_{2,\infty}
\ln^2\frac{\hbar\omega_{\text{L}}}{\varepsilon_{\text{F}}}
\Delta_2
- \alpha_{3,\infty} \Delta_3~,
\end{multlined}
\end{gather}
\end{subequations}
where
\begin{gather}
\alpha_{1,i} =
\begin{dcases}
\frac{\alpha_{\text{ee}}}{\PI\epsilon_i}
\ln\frac{4\epsilon_i}{\E N_{\text{f}}\alpha_{\text{ee}}}~,
& \frac{N_{\text{f}}\alpha_{\text{ee}}}{2\epsilon_i}\ll1~, \\
\frac{1}{2N_{\text{f}}}~,
& \frac{N_{\text{f}}\alpha_{\text{ee}}}{2\epsilon_i}\gg1~,
\end{dcases}
\\
\alpha_{2,i} = \frac{\alpha_{\text{ee}}}{2\pi\epsilon_i}~,
\quad
\alpha_{3,i} = \frac{6\ln2}{\pi}\,
\frac{\alpha_{\text{ee}}}{2\pi\epsilon_i}~,
\end{gather}
and the electron-electron coupling constant is defined (for 2D MDFs) in the usual manner, i.e.~$\alpha_{\text{ee}}\equiv e^2/\hbar v$.
For $T=T_{\text{c}}^-$, the system has a non-zero solution,
i.e., its determinant is zero.
This provides a quadratic equation for $\ln T_{\text{c}}$.
Analogous calculations can be performed in the case
$k_{\text{B}}T\ll\varepsilon_{\text{F}}\ll\hbar\omega_{\text{L}}$,
with the approximate gap
\begin{equation}
\Delta(\I\omega) =
\begin{cases}
\Delta_1~, & \lvert\hbar\omega\rvert < \varepsilon_{\text{F}}~, \\
\Delta_2~, & \varepsilon_{\text{F}} < \lvert\hbar\omega\rvert < \hbar\omega_{\text{L}}~, \\
\Delta_3~, & \lvert\hbar\omega\rvert < \hbar\omega_{\text{L}}~,
\end{cases}
\end{equation}
and the linear system is
\begin{subequations}
\begin{gather}
\begin{multlined}[b]
\Delta_1
= - \alpha_{1,0}\ln\frac{\varepsilon_{\text{F}}}{k_{\text{B}} T} \Delta_1
\\
+ \alpha_{3,0}\biggr(
\frac{\varepsilon_{\text{F}}}{\hbar\omega_{\text{L}}}-1\biggl) \Delta_2
- \alpha_{3,\infty}\frac{\varepsilon_{\text{F}}}{\hbar\omega_{\text{L}}} \Delta_3~,
\end{multlined}
\\
\begin{multlined}[b]
\Delta_2
= -\alpha_{2,0}
\ln^2\frac{k_{\text{B}} T}{\varepsilon_{\text{F}}} \Delta_1
\\
+ \alpha_{3,0}\biggr(
\frac{\varepsilon_{\text{F}}}{\hbar\omega_{\text{L}}}-1\biggl) \Delta_2
- \alpha_{3,\infty}\frac{\varepsilon_{\text{F}}}{\hbar\omega_{\text{L}}} \Delta_3~,
\end{multlined}
\\
\begin{multlined}[b]
\Delta_3
= -\alpha_{2,\infty}
\ln^2\frac{k_{\text{B}} T}{\varepsilon_{\text{F}}} \Delta_1
\\
+ \alpha_{3,\infty}\biggr(
\frac{\varepsilon_{\text{F}}}{\hbar\omega_{\text{L}}}-1\biggl) \Delta_2
- \alpha_{3,\infty}\frac{\varepsilon_{\text{F}}}{\hbar\omega_{\text{L}}} \Delta_3~.
\end{multlined}
\end{gather}
\end{subequations}

The analytical solution for $T_{\text{c}}$ does not give much insight,
having an involved algebraic expression.
Therefore, we now explicitly consider some limiting cases.
In the limit $\hbar\omega_{\text{L}}\ll\varepsilon_{\text{F}}$,
\begin{multline}
k_{\text{B}}T_{\text{c}}
\sim \varepsilon_{\text{F}} \exp\Biggl[
- \frac{\alpha_{1,\infty}(1+\alpha_{3,\infty})}{
2\alpha_{2,\infty}\alpha_{3,\infty}}
\\
\times
\Biggl(
1 + \sqrt{1 + \frac{4\alpha_{2,\infty}\alpha_{3,\infty}}{
(\alpha_{1,\infty})^2(1+\alpha_{3,\infty})}}\,
\Biggr)
\Biggr]~.
\end{multline}
In the limit $\hbar\omega_{\text{L}}\sim\varepsilon_{\text{F}}$,
\begin{multline}
k_{\text{B}}T_{\text{c}}
\sim \varepsilon_{\text{F}} \exp\Biggl[
- \frac{\alpha_{1,0}(1+\alpha_{3,\infty})}{
2\alpha_{2,\infty}\alpha_{3,\infty}}
\\
\times
\Biggl(
1 + \sqrt{1 + \frac{4\alpha_{2,\infty}\alpha_{3,\infty}}{
(\alpha_{1,0})^2(1+\alpha_{3,\infty})}}\,
\Biggr)
\Biggr]~.
\end{multline}
In the limit $\hbar\omega_{\text{L}}\gg\varepsilon_{\text{F}}$,
\begin{multline}
k_{\text{B}}T_{\text{c}}
\sim \varepsilon_{\text{F}} \exp\Biggl[
- \frac{\alpha_{1,0}(1+\alpha_{3,0})}{
2\alpha_{2,0}\alpha_{3,0}}
\\
\times
\Biggl(
1 + \sqrt{1 + \frac{4\alpha_{2,0}\alpha_{3,0}}{
(\alpha_{1,0})^2(1+\alpha_{3,0})}}\,
\Biggr)
\Biggr]~.
\end{multline}
Thus, in general, the critical temperature increases by decreasing the MDF
velocity $v$, i.e.~by increasing the coupling
constant $\alpha_{\text{ee}}$.
Furthermore, the critical temperature increases by decreasing the
permittivity $\epsilon_\infty$ (i.e.~reducing screening) and the energy of the longitudinal optical phonon with respect to the relevant electronic energy scales (i.e.~Fermi energy/plasmon energy, see comment below).

\begin{figure}
\centering
\includegraphics{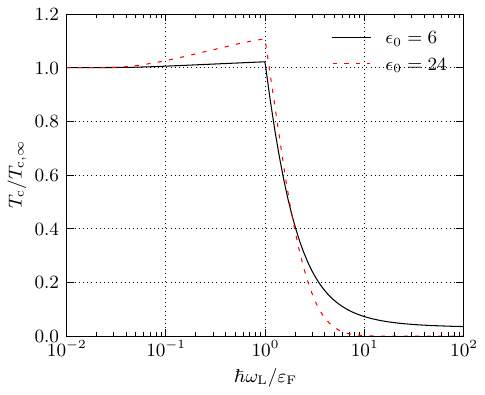}
\caption{%
The approximate, analytically obtained, critical temperature $T_{\text{c}}$ of the 2D system of MDFs (measured in units of $T_{\text{c},\infty}$) is plotted as a function of the energy
of the longitudinal optical phonon mode $\hbar\omega_{\text{L}}$ (measured in units of $\varepsilon_{\text{F}}$).
The system is sketched in the inset of Fig.~\ref{fig:one}.
The dielectric has a high-frequency permittivity $\epsilon_\infty=3$,
while the static permittivity $\epsilon_0$ is reported in the legend.
The parameters are:
$d_{\text{b}}=d_{\text{t}}\to\infty$,
$n = \SI{e11}{cm^{-2}}$,
$v=v_0/2$ ($\theta=2.2^\circ$ in TBG).
}\label{fig:tcphononapprox}
\end{figure}

\begin{figure}
\centering
\includegraphics{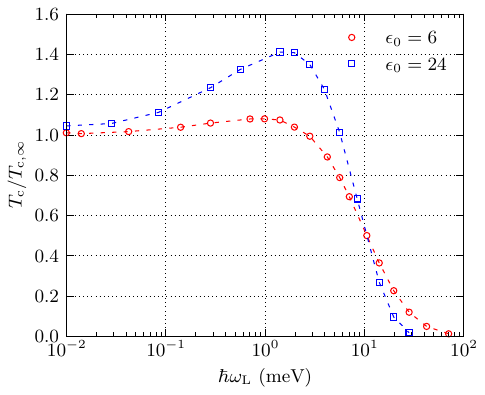}
\caption{%
Plot the ratio $T_{\text{c}}/T_{\text{c},\infty}$
as a function of the longitudinal optical phonon frequency $\hbar\omega_{\text{L}}$.
The system is sketched in the inset of Fig.~\ref{fig:one}.
The dielectric has a high-frequency permittivity $\epsilon_\infty=3$,
while the static permittivity $\epsilon_0$ is reported in the legend.
The parameters are:
$d_{\text{b}}=d_{\text{t}}\to\infty$,
$n = \SI{e11}{cm^{-2}}$,
$v=v_0/2$ ($\theta=2.2^\circ$ in TBG).
}\label{fig:tcphonon}
\end{figure}

In Fig.~\ref{fig:tcphononapprox} we report the critical temperature as a function of the phonon energy, as obtained by using the piecewise-constant approximation of the gap equation. We clearly see that our analytical approximation reproduces, at least qualitatively, the fully numerical result reported in Fig.~\ref{fig:one} of the main text.
In Fig.~\ref{fig:tcphonon}, we report the numerical results of Fig.~\ref{fig:one} of the main text normalized by the value $T_{\text{c},\infty}$.

We note that in this approximate treatment of the gap equation, the critical temperature is defined up to a constant, with logarithmic accuracy. Also, in our approximate analytical treatment we have completely neglect the dispersion of the 2D plasmon mode $\hbar \omega_{\rm p}(q)$, considering the Fermi energy $\varepsilon_{\rm F}$ as the characteristic scale of this excitation.
This crude approximation greatly overestimates the absolute value of the critical temperature $T_{\rm c}$ but a partial cancellation of errors occur in the ratio $T_{\rm c}/T_{{\rm c}, \infty}$. (The scale of the critical temperature is determined by the energy scale of the bosonic glue, which is effectively much smaller, because the plasmon spectral weight is larger at small energies.)
However, the qualitative behavior is well captured and can offer a useful insight into the parameters which allow the onset of superconductivity.


\section{The role of metal gates}

\begin{figure}
\centering
\includegraphics{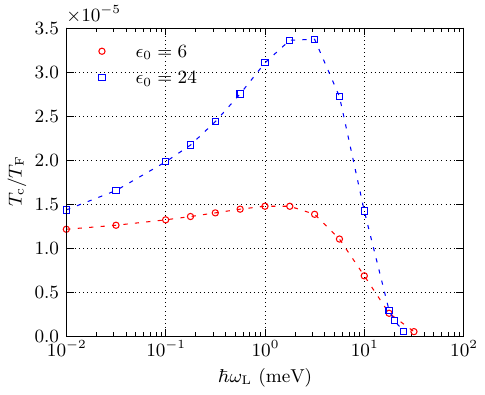}
\caption{%
Plot the ratio $T_{\text{c}}/T_{\text{F}}$
as a function of the longitudinal optical phonon frequency $\hbar\omega_{\text{L}}$.
We consider a device where the 2DES is encapsulated between two dielectric slabs of thickness $d_{\text{t}}$ and $d_{\text{b}}$, with the same permittivity $\epsilon(\omega)$, and metal top and bottom gates above and below, respectively.
The dielectric has a high-frequency permittivity $\epsilon_\infty=3$,
while the static permittivity $\epsilon_0$ is reported in the legend.
The parameters are:
$d_{\text{t}}=d_{\text{b}}=\SI{300}{nm}$,
$n = \SI{e11}{cm^{-2}}$,
$v=v_0/2$ ($\theta=2.2^\circ$ in TBG).
}\label{fig:tcphonond}
\end{figure}

In experiments,
the carrier density is controlled by field effect with metal gates.
Here, we consider the effect of such gates (neglected in the main text).
We consider a device where the 2DES is encapsulated between two dielectric slabs of thickness $d_{\text{t}}$ and $d_{\text{b}}$, with the same permittivity $\epsilon(\omega)$, and metal top and bottom gates above and below, respectively.
In this case, the electron-electron interaction is
\begin{equation}
V(q,\omega) = \frac{2\pi e^2}{\epsilon(\omega)q} \,
\frac{2\sinh(qd_{\text{t}})\sinh(qd_{\text{b}})}{\sinh(q(d_{\text{t}}+d_{\text{b}}))}.
\end{equation}
The metal gate screens the long range electron-electron interaction
(on a scale $d_{\text{b}}$, $d_{\text{t}}$),
effectively decreasing the coupling.
Since we consider an electronic mechanism of superconductivity,
the critical temperature decreases with decreasing gate distance.
Consider the critical temperature for the plasmon mechanism
$T_{\text{c},\infty}/T_{\text{F}}=\num{3.3e-4}$,
which is obtained by the numerical solution of the gap equation
for $v=v_0/2$, $n=\SI{e11}{cm^{-2}}$, $N_{\text{f}}=8$,
$d_{\text{b}}=d_{\text{t}}\to\infty$
(as in Fig.~\ref{fig:one} of the main text).
If the gate distance is finite, $d_{\text{b}}=d_{\text{t}}=\SI{300}{nm}$,
the critical temperature is reduce to
$T_{\text{c},\infty}/T_{\text{F}}=\num{1.1e-5}$.
If the distance is further reduced to \SI{100}{nm},
the critical temperature becomes
$T_{\text{c},\infty}/T_{\text{F}}=\num{2.1e-8}$.
In Fig.~\ref{fig:tcphonond}, we show the critical temperature $T_{\text{c}}/T_{\text{F}}$
as a function of the longitudinal optical phonon energy $\hbar\omega_{\text{L}}$ of the dielectric,
analogously to Fig.~\ref{fig:one} of the main text
but with top and bottom gates at distance $d_{\text{b}}=d_{\text{t}}=\SI{300}{nm}$.
We note that, even in this gated case,
there is an enhancement of the critical temperature
when the phonon energy is resonant with the plasmon energy,
qualitatively analogous to the ungated limit $d_{\text{b}}=d_{\text{t}}\to\infty$,
reported in Fig.~\ref{fig:one} of the main text.

\begin{figure}
\centering
\begin{overpic}[percent]{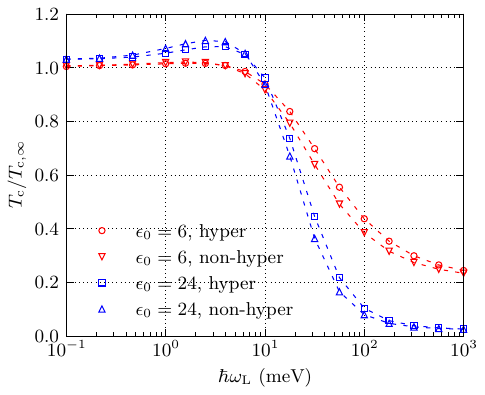}
\put(0,80){\makebox(0,0)[tl]{(a)}}
\end{overpic}
\begin{overpic}[percent]{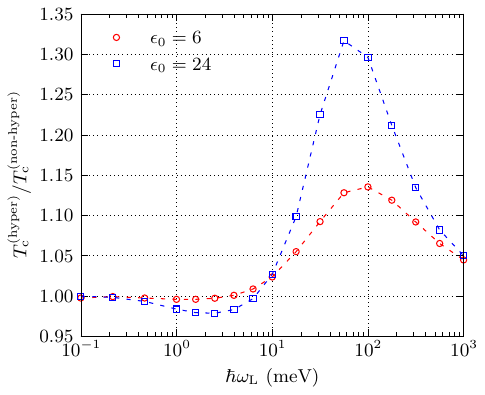}
\put(0,80){\makebox(0,0)[tl]{(b)}}
\end{overpic}
\caption{The role of hyperbolicity.
(a) The ratio $T_{\text{c}}/T_{\text{c},\infty}$ is plotted as a function of $\hbar\omega_{\text{L}}$ for a 2DES encapsulated between two dielectric slabs of finite thicknesses, $d_{\text{b}}=d_{\text{t}}=\SI{100}{nm}$.
Fermi velocity in the 2DES and carrier density are as in the previous plots, i.e.~$v=v_0/2$ ($\theta=2.2^\circ$) and $n = \SI{e11}{cm^{-2}}$. Parameters in the hyperbolic case (red circles and blue squares) are fixed as described in the text---see Eq.~(\ref{eq:simple_permittivity_hyperbolic}).
(b) Ratio between the critical temperature
$T_{\text{c}}^{\text{(hyper)}}$ in the hyperbolic case and the one in the non-hyperbolic case, $T_{\text{c}}^{\text{(non-hyper)}}$.
All parameters are the same as in panel (a).
}\label{fig:tchyper}
\end{figure}

\section{The role of hyperbolicity}
We conclude by investigating whether {\it hyperbolic} dielectric slabs help boosting $T_{\rm c}$ with respect to the non-hyperbolic case discussed so far. To this end, one needs to properly introduce a uniaxial electrical permittivity tensor $\hat{\bm \epsilon}(\omega) = (\epsilon_x(\omega), \epsilon_x(\omega), \epsilon_z(\omega))$ into the theory of the dressed propagator. This leads to a different function $V(q,\omega)$ with respect to the one reported in Eqs.~(\ref{eq:dressed_propagator_non_hyperbolic})-(\ref{eq:form_factor_non_hyperbolic}) of the main text, which we dub $V_{\rm hyper}(q,\omega)$.
The analytical expression of the propagator $V_{\rm hyper}(q,\omega)$ for a 2DES encapsulated between two hyperbolic dielectrics of different thicknesses $d_{\rm t}$ and $d_{\rm b}$---as in the inset in Fig.~\ref{fig:one}---is
\begin{equation}\label{eq:propagator-hyper}
V_{\rm hyper}(q,\omega) = \frac{2\pi e^2}{
q\epsilon_z(\omega)r(\omega)}F_{\rm hyper}(q,\omega),
\end{equation}
where 
\begin{equation}
r(\omega) \equiv \sqrt{\frac{\epsilon_{x}(\omega)}{\epsilon_{z}(\omega)}}
\end{equation}
and the hyperbolic form factor $F_{\rm hyper}(q,\omega)$ reads as following:
\begin{multline}
F_{\rm hyper}(q,\omega)  \equiv \biggl[ \frac{1}{2}\,
\frac{1 + \epsilon_z(\omega)r(\omega)\tanh(r(\omega)d_{\text{t}}q)}{
\tanh(r(\omega)d_{\text{t}}q) + \epsilon_z(\omega)r(\omega)}
\\
+  \frac{1}{2}\,
\frac{1 + \epsilon_z(\omega)r(\omega)\tanh(r(\omega)d_{\text{b}}q)}{
\tanh(r(\omega)d_{\text{b}}q) + \epsilon_z(\omega)r(\omega)}
\biggr]^{-1}.
\end{multline}
Hyperbolic behavior~\cite{Sun_ACS_2014} implies that the quantity $\epsilon_x(\omega)/\epsilon_z(\omega)$ takes negative values in the lower (upper) reststrahlen band, which is defined by the inequality $\omega^{\rm T}_{z} < \omega < \omega^{\rm L}_{z}$ ($\omega^{\rm T}_{x} < \omega < \omega^{\rm L}_{x}$). Inside the reststrahlen bands, the dressed propagator $V_{\rm hyper}(q, \omega)$ displays poles, which physically correspond to standing hyperbolic phonon polaritons (SHPPs)~\cite{Basov_Science_2016,Low_NatMater_2017,Zhang_Nature_2021,Basov_Nanophotonics_2021,Plantey_ACS_2021}. These modes have been measured in many natural hyperbolic materials, including hexagonal boron nitride (hBN)~\cite{dai_science_2014,caldwell_naturecommun_2014,li_NatCommun_2015,dai_NatCommun_2015} and many others, and span a broad range of energies, from the mid-infrared to the Terahertz.

We now move on to analyze the role of hyperbolicity on the critical temperature $T_{\rm c}$. To this end, we introduce parametrizations of the in-plane ($\alpha=x$) and out-of-plane ($\alpha=z$) permittivities as following
\begin{equation}\label{eq:simple_permittivity_hyperbolic}
\frac{1}{\epsilon_\alpha(\omega)}
= \frac{1}{\epsilon_{\infty, \alpha}}
+ \biggl(\frac{1}{\epsilon_{\infty, \alpha}} - \frac{1}{\epsilon_{0, \alpha}}\biggr)
\frac{\omega_{{\rm L}, \alpha}^2}{\omega^2 - \omega_{{\rm L}, \alpha}^2}~.
\end{equation}
Given the amplitude of parameter space, we take, for the sake of simplicity,  $\epsilon_{\infty,x}=\epsilon_{\infty,z} \equiv \epsilon_\infty$, $\epsilon_{0,x}=\epsilon_{0,z} \equiv \epsilon_0$, $\omega_{\text{L},x}=\omega_{\text{L}}$, and $\omega_{\text{L},z}=\omega_{\text{L}}/\delta$ with $\delta > \sqrt{\epsilon_0 / \epsilon_\infty}$. (This condition on $\delta$ ensures that, while changing the phonon oscillator strength, the upper and lower reststrahlen bands do not overlap.)
The quantities $\omega_{\text{T},x}$ and $\omega_{\text{T},z}$ are then determined by the LST relation (\ref{eq:LST_relation}), i.e.~$\omega^2_{{\rm L}, \alpha}/\omega^2_{{\rm T}, \alpha}
=\epsilon_{0, \alpha}/\epsilon_{\infty, \alpha} = \epsilon_0/\epsilon_\infty$. In our numerical calculations below, we fix $\epsilon_\infty=3$, as in the non-hyperbolic case studied so far, and we change $\epsilon_0$ to change the phonon oscillator strength. Finally, we take $\delta=3$. In this manner, we can compare results for the non-hyperbolic case with results for the hyperbolic case in the same plot. 

A summary of our main results is reported in Fig.~\ref{fig:tchyper}, where we analyze superconducting critical temperatures in a 2DES encapsulated between two dielectric slabs of finite thickess $d_{\rm t} = d_{\rm b} = 100~{\rm nm}$. We remind the reader that in the main text, we have studied the case of two semi-infinite dielectric slabs, i.e.~$d_{\rm t} = d_{\rm b} \to \infty$. Reducing the thickness to a finite value plays a dramatic role in the absolute magnitude of $T_{{\rm c}, \infty}$. In the case of Fig.~\ref{fig:one}, we have $T_{{\rm c}, \infty} =\SI{0.05}{K}$ while in the case of Fig.~\ref{fig:tchyper} we find $T_{{\rm c}, \infty} =\SI{0.64}{K}$. We also note that reducing the Fermi velocity from $v = v_0/2$ ($\theta=2.2^\circ$) to $v = v_0/5$ ($\theta=1.46^\circ$) yields values of $T_{{\rm c}, \infty}$ easily exceeding $\SI{1}{K}$. Looking at panels (a) and (b) in Fig.~\ref{fig:tchyper}, we conclude that hyperbolicity plays a positive role (but not a pivotal one) since the ratio between the critical temperature in the hyperbolic case, i.e.~$T^{({\rm hyper})}_{\rm c}$, and the one in the non-hyperbolic case, i.e.~$T^{({\rm non-hyper})}_{\rm c}$, is typically larger than unity in  the explored range of values of $\hbar\omega_{\rm L}$.


\begin{thebibliography}{77}
\bibitem{Hyperbolicity}
The general case---see inset in Fig.~\ref{fig:one}---in which top and bottom slabs have different thicknesses, $d_{\rm t}$ and $d_{\rm b}$, and are described by a uniaxial dielectric tensor $\hat{\bm \epsilon}(\omega)=(\epsilon_x(\omega), \epsilon_x(\omega), \epsilon_z(\omega))$---which is necessary for example in the case of hyperbolic dielectrics such as hBN---is discussed in Ref.~\cite{SM}.
\end{thebibliography}
\end{document}